\documentclass[conference]{IEEEtran}
\IEEEoverridecommandlockouts

\usepackage{xcolor}
\usepackage[pdftex]{graphicx}
\graphicspath{{./}}
\usepackage{wrapfig}

\usepackage[colorlinks=true,linkcolor=blue,citecolor=blue,hyperfootnotes]{hyperref}
\usepackage{cite,url}

\usepackage{expdlist}

\usepackage{amsmath,amssymb,amsmath,amsfonts}

\DeclareMathOperator{\ev}{\mathsf{ev}}
\DeclareMathOperator{\id}{\mathsf{id}}
\DeclareMathOperator{\creator}{\mathsf{creator}}
\DeclareMathOperator{\type}{\mathsf{type}}
\DeclareMathOperator{\own}{\mathsf{own}}
\DeclareMathOperator{\dsc}{\mathsf{dsc}}

\DeclareMathOperator{\tm}{\mathsf{time}}
\DeclareMathOperator{\Tx}{\mathsf{Tx}}
\DeclareMathOperator{\Hash}{\mathsf{Hash}}
\DeclareMathOperator{\nonce}{\mathsf{nonce}}

\DeclareMathOperator{\CreateEvidence}{\mathsf{CreateEvidence}}
\DeclareMathOperator{\TransferOwnership}{\mathsf{TransferOwnership}}
\DeclareMathOperator{\EraseEvidence}{\mathsf{EraseEvidence}}
\DeclareMathOperator{\GetEvidence}{\mathsf{GetEvidence}}

\def\BibTeX{{\rm B\kern-.05em{\sc i\kern-.025em b}\kern-.08em
 T\kern-.1667em\lower.7ex\hbox{E}\kern-.125emX}}

\begin{document}

\thispagestyle{empty}

\begin{figure*}[!t]\large
This paper is a preprint; it has been accepted for publication in 2019 IEEE Conference on Network Softwarization (IEEE NetSoft), 24--28 June 2019, Paris, France.
\medskip

{\bf IEEE copyright notice}
\smallskip

\copyright\ 2019 IEEE. Personal use of this material is permitted. Permission from IEEE must be obtained for all other uses, in any current or future media, including reprinting/republishing this material for advertising or promotional purposes, creating new collective works, for resale or redistribution to servers or lists, or reuse of any copyrighted component of this work in other works.
\vspace*{300pt}

\mbox{~}
\end{figure*}

\newpage

\title{Blockchain Solutions for Forensic Evidence Preservation in IoT Environments%
\thanks{%
\protect\begin{wrapfigure}[3]{l}{.9cm}%
\protect\raisebox{-12.5pt}[0pt][0pt]{\protect\includegraphics[height=.8cm]{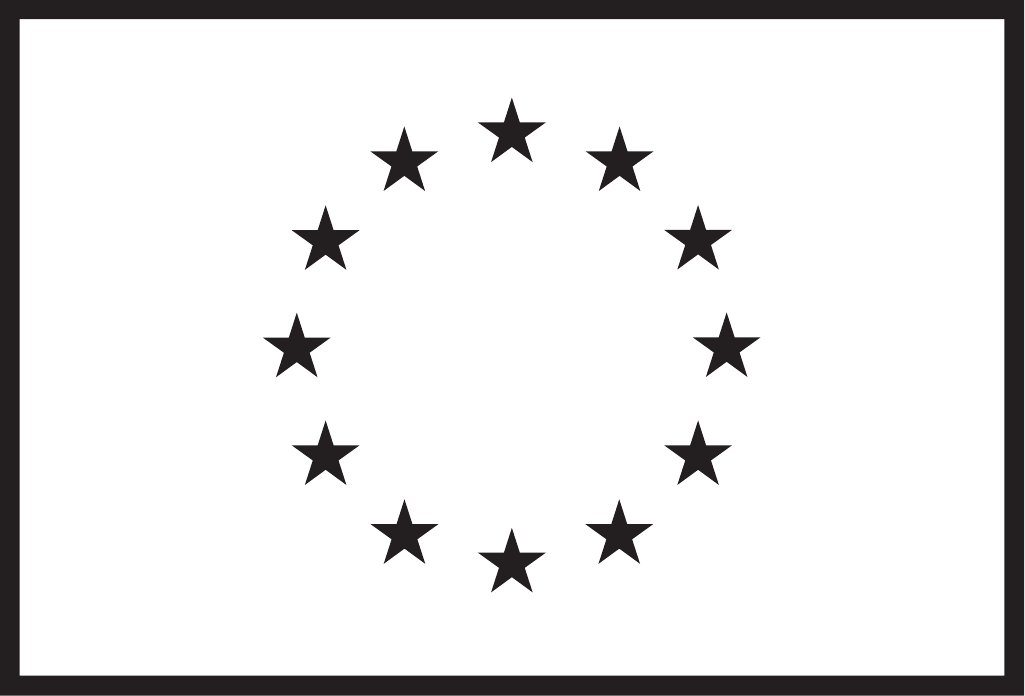}}%
\protect\end{wrapfigure}%
This project has received funding from the European Union's Horizon 2020 research and innovation programme under grant agreement no. 786698. The work reflects only the authors' view and the Agency is not responsible for any use that may be made of the information it contains.}}

\author{\IEEEauthorblockN{%
Sotirios Brotsis\IEEEauthorrefmark{1},
Nicholas Kolokotronis\IEEEauthorrefmark{1},
Konstantinos Limniotis\IEEEauthorrefmark{1},
Stavros Shiaeles\IEEEauthorrefmark{2},
Dimitris Kavallieros\IEEEauthorrefmark{3},\\
Emanuele Bellini\IEEEauthorrefmark{4},
and Cl\'{e}ment Pavu\'{e}\IEEEauthorrefmark{5}
\vspace*{4pt}}

\IEEEauthorblockA{\IEEEauthorrefmark{1}University of Peloponnese, Greece. Email: brotsis@uop.gr, nkolok@uop.gr, klimn@uop.gr}

\IEEEauthorblockA{\IEEEauthorrefmark{2}Plymouth University, UK. Email: stavros.shiaeles@plymouth.ac.uk}

\IEEEauthorblockA{\IEEEauthorrefmark{3}Center for Security Studies, Greece. Email: d.kavallieros@kemea-research.gr}


\IEEEauthorblockA{\IEEEauthorrefmark{4}Mathema s.r.l., Italy; Khalifa University, UAE. Email: emanuele.bellini@mathema.com}

\IEEEauthorblockA{\IEEEauthorrefmark{5}Scorechain S.A., Luxembourg. Email: clement.pavue@scorechain.com}
}

\maketitle


\begin{abstract}
The technological evolution brought by the {\em Internet of things} (IoT) comes with new forms of cyber-attacks exploiting the complexity and heterogeneity of IoT networks, as well as, the existence of many vulnerabilities in IoT devices. The detection of compromised devices, as well as the collection and preservation of evidence regarding alleged malicious behavior in IoT networks emerge as a areas of high priority. This paper presents a blockchain-based solution, which is designed for the smart home domain, dealing with the collection and preservation of digital forensic evidence. The system utilizes a private forensic evidence database, where the captured evidence is stored, along with a permissioned blockchain that allows providing security services like integrity, authentication, and non-repudiation, so that the evidence can be used in a court of law. The blockchain stores evidences' metadata, which are critical for providing the aforementioned services, and interacts via smart contracts with the different entities involved in an investigation process, including Internet service providers, law enforcement agencies and prosecutors. A high-level architecture of the blockchain-based solution is presented that allows tackling the unique challenges posed by the need for digitally handling forensic evidence collected from IoT networks.
\end{abstract}



\section{Introduction}
\label{sec.intro}

The {\em Internet of things} (IoT) ecosystem is comprised of a vast number of interconnected devices that collect, process, generate, and share huge amounts of (possibly sensitive and critical) information \cite{Cheng2018}. To a large extent, these devices are highly resource-constrained, like sensors and legacy embedded systems, therefore devoting most of their computational power and storage/memory capacity to delivering their core functionality. Strong security controls that are typically found in today's personal computers cannot be adopted, since they are more resource-demanding, hence leading to the usage of lightweight and often insecure protection mechanisms (if any) for the data stored or transmitted. This fact, if combined with the complexity and heterogeneity of IoT networks that make the design and provisioning of security solutions a challenging task \cite{Zhao2013}, allows cyber-attackers to easily compromise them and use them as the means for launching other advanced attacks, such as the {\em distributed denial of service} (DDoS) attack against Dyn that was attributed to Mirai malware \cite{Kolias2017}.

The collection of forensic evidence from the attacked IoT devices and networks, along with their storage, preservation, and analysis constitute major challenges \cite{MacDermott2018}, primarily due to the fact that IoT devices are designed to work autonomously and in many cases, there is no reliable method to assemble residual evidence \cite{Orazio2017}. The utilization of {\em intrusion detection systems} (IDS) in the collection process is important towards identifying cyber-criminals and preventing future occurrences of attacks \cite{Fung2008}. In an IoT environment, the identification of a crime scene's boundaries and its preservation are quite hard to accomplish while interactions continuously occur at real-time. Since the majority of IoT devices are sensors and monitors that record user's personal information, privacy is an important issue to consider in a digital forensics investigation.

This paper aims at addressing the challenges in the forensic evidence collection, preservation and investigation process, for IoT environments in the smart home domain, by exploiting the advanced intrusion detection and {\em distributed ledger technology} (DLT) solutions that are being developed in the context of the Cyber-Trust project. More precisely, a number of mechanisms installed at a smart home's gateway, like profiling, monitoring, and anomaly detection, allow to monitor the state and behavior of IoT devices, significantly enhancing the detection of known threats and zero-day vulnerabilities, as well as to immediately collect forensic evidence for detected malicious interactions. The collected data are stored at the {\em evidence database} (evDB), hosted by the {\em Internet service provider} (ISP), along with the metadata needed in order to allow the correlation and further investigation of an attack's generated events. The metadata are published on a blockchain, which is maintained by the ISPs, maintaining the chronological ordering of attacks' evidences at a global scale, thus providing the means to {\em law enforcement agencies} (LEA) to effectively trace back an attack to its source. The proposed solution, referred to as {\em Cyber-Trust blockchain} (CTB), allows the entities involved in the investigation process, such as LEAs and prosecutors, to access and handle the digital evidence, therefore realizing the {\em chain-of-custody} (CoC) by recording and preserving the chronological history of handling the digital evidence. The CTB solution relies on HyperLedger Fabric and constitutes a permissioned blockchain in order to meet privacy requirements.

The remainder of the paper is structured as follows. Section \ref{sec.related} presents the current state-of-the-art and related work, while the forensic evidence collection process is described in Section \ref{sec.ev.store}. Section \ref{sec.ctb} provides the architecture of the CTB solution whereas concluding remarks are given in Section \ref{sec.concl}.


\section{Background and related work}
\label{sec.related}

This section presents the current state-of-the-art in the areas of intrusion detection and forensic evidence collection for IoT environments, along with the blockchain solutions that have been proposed.


\subsection{IoT intrusion detection}
\label{sec.ids}

Intrusion detection systems typically utilize signature-based and anomaly-based techniques for identifying possible threats in a network, where the latter relies on the monitoring of a network's devices for any abnormal behavioral patterns \cite{Fung2008}. In order to detect compromised IoT devices, the framework that is proposed by Nguyen, {\em et al.} autonomously identifies anomalies in an IoT network \cite{Nguyen2018}; this is achieved by employing a self-learning framework to classify devices according to their types and generate normal profiles that are subsequently used for the detection of deviations. A privacy-preserving architecture, called Siotome, was proposed in \cite{Haddadi2017} to provide security in smart home environments against distributed network attacks by malicious IoT devices; the system is able to monitor, detect and analyze IoT-based threats, but also to provide an effective defense framework by utilizing machine learning methods to establish optimal operational configurations.

Smart phones are a particular type of devices within a smart home environment, since they are mostly used for personal and sensitive tasks, thus becoming extremely beneficial and easy targets for adversaries. Smart phones, which are vulnerable to attacks (e.g. viruses, Trojans, worms, etc.) common in personal computers, but they lack the capabilities to execute highly advanced algorithms for detecting malicious activities. Due to this fact, IDS solutions that are often proposed to regularly perform in-depth analysis and observe any misbehavior are either cloud-based \cite{Houmansadr2011}, or are performed remotely at a central server \cite{Schmidt2009}, allowing optimal actions to be taken for thwarting the attack in both architectures.


\subsection{IoT forensics}
\label{sec.evidence}

The wide adoption of smart devices, which can provide a wealth of forensic evidence on malicious activities during an investigation process, necessitated the advancement of tools and techniques for collecting residual evidences. A forensics edge management system for the smart home environment was introduced in \cite{Oriwoh201342} to gather digital evidence and deal with any security issues; it provides intelligence, flexibility, automated detection, and advanced data logging capabilities. The authors in \cite{Goudbeek2018} proposed a forensic investigation architecture to ensure the collection, preservation and storage of digital evidences, while they validated their approach in a real-world smart home environment. Focusing on a smart home's IoT devices, a physical analyzer called {\em universal forensic extraction device} has been proposed for conducting forensic investigation on smart phones \cite{Faheem2014}, which has been tested on Android devices. A comparative analysis of digital forensics tools for Android smart phones was carried out in \cite{Raji2018AnalyzingDF}, where it was illustrated that the choice of the tool to be used plays a crucial role in the quality of the forensic evidence that is extracted from the devices. In contrast to \cite{Raji2018AnalyzingDF}, a method for acquiring forensic evidence from Android smart phones without using specialized commercial forensics tools, i.e. by only relying on open source software, was proposed in \cite{Andriotis2012}. In all the above works, it was shown that the collection of information from smart phones so that it can be used as evidence in a court of law still remains a challenging task.


\subsection{Blockchain solutions}
\label{sec.bc.sol}

Blockchain solutions have recently been proposed for both intrusion detection and forensic evidence applications, since in both cases blockchain can solve issues pertaining to trust, integrity, transparency, accountability, and secure data sharing. Addressing the issue of trust management, Alexopoulos, {\em et al.} \cite{Alexopoulos2018} applied blockchain in collaborative intrusion detection networks to deal with insider threats but also enhance the security of the information shared among the participating IDS nodes. More precisely, the authors proposed to store the generated (raw) alerts of the network as transactions in a permissioned blockchain. Meng, {\em et al.} \cite{Meng2018} in addition to the dimension of trust between the IDS nodes, refer to issues that pertain to privacy when collaborating nodes belong to different trust domains, as shared data may have sensitive information linked to individuals or organizations, e.g., IP addresses and packet payloads. Methods for exchanging encrypted content, or only hashed data rather than raw, are considered.

In forensic investigations, it is important that the evidence is not modified while passing from one entity to another. The blockchain can be used in order to certify the authenticity and legitimacy of the procedures used to gather, store and transfer digital evidence, as well as, to provide a comprehensive view of all the interactions in the CoC. In a blockchain-based CoC, it is crucial to assure that members, having read/write access to the distributed ledger, are authenticated and the evidences are verified via a consensus algorithm. Towards that direction, Lone, {\em et al.} propose a private blockchain that can be used in digital forensics to ensure the integrity of evidences \cite{Lone2019}; the authors also aim at recording the actions taken by each entity when interacting with the evidence. On the other hand, Probe-IoT uses a blockchain to discover criminal events, which can be used as evidence, by collecting interactions between IoT devices and verify their authenticity \cite{Hossain2018}. 


\section{Forensic evidence collection architecture}
\label{sec.ev.store}

The primary goal of Cyber-Trust in the smart home domain, or in general in {\em small office\,/\,home office} (SOHO) network, is to accurately detect the local network's compromised and/or infected IoT devices to apply the appropriate countermeasures, e.g. to isolate the devices from the rest of the network and to proceed with the application of proper remediation measures. The intrusion detection mechanisms that are being employed are operating both at the device- and network-level to facilitate the collection and subsequent correlation of forensic evidence from various independent sources.

To combat cyber-attacks and assist the evidence collection, IoT devices' critical information is recorded on the blockchain so that it can be later queried when e.g. a verification of proper functioning is needed, or parts of the system's software have to be updated or patched reliably. This implies that properties, like a device's firmware, configuration files, etc. are registered into the Cyber-Trust blockchain, at the beginning of system's operation, and verified if needed against a history of previously valid states, in order to ensure that they have not been tampered with. This approach fits well within the practices of software distributors that publish hashes of software binaries to allow verifying their authenticity.


\subsection{Adversarial model}
\label{sec.adversaries}

The adversary is a typical IoT malware botnet that actively scans for vulnerable Linux-based IoT devices in the SOHO network, like smart watches, home surveillance systems, smart phones, etc., and infects the discovered vulnerable devices by uploading and executing malware code of an unknown bot on the compromised devices; once infected, the IoT devices may take a variety of malicious actions. Typically, the phases of a botnet, prior to performing attacks in a coordinated manner, are the following.

\subsubsection{Propagation}
If having been infected with malware, a smart home's device updates its configuration and downloads further exploits. The bot replicates itself in the SOHO network using telnet/FTP/SSH default credentials and attacks nearby devices with firmware vulnerabilities.

\subsubsection{Rallying}
The bot contacts a {\em command \& control} (C\&C) server, queries for instructions, and also downloads the main configuration files. The bot and the bot-master share a seeded pseudorandom generator that computes the domain names.

\subsubsection{Interaction}
Bot-masters use a pull approach, in which the bot should initiate contact with the C\&C server, and then poll for updates regularly. Obfuscation techniques are used, by hiding communications in regular web traffic, hence allowing perimeter controls to be bypassed.

As seen from above, the bot is listening for commands via the HTTP and HTTPS protocols (utilizing ports 80 and 443) and is assumed to execute three types of attacks, namely {\em man-in-the-middle} (MiTM), DDoS, and spamming. 

\begin{figure}[t]
\centering
\includegraphics[width=\linewidth]{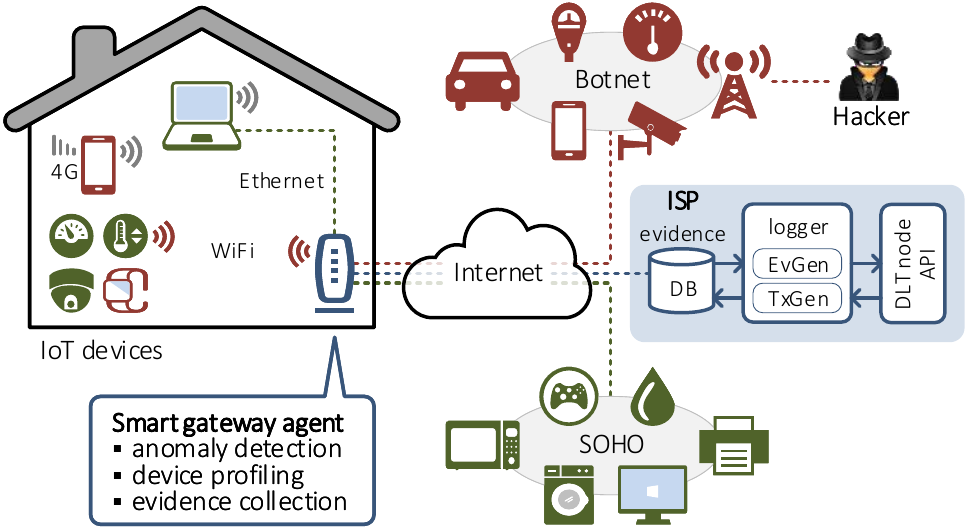}
\caption{An overview of Cyber-Trust's forensic evidence collection process; it is assumed that the red-colored devices in the smart home have been attacked and this is detected by the SGA that collects the evidence.}
\label{fig.collection}
\end{figure}


\subsection{Architectural elements}
\label{sec.smart.home}

In the sequel, we describe the high-level design of the smart home environment's security elements, as illustrated in Fig. \ref{fig.collection}. The {\em smart gateway agent} (SGA) is the core component that is responsible for the smart home's network security by utilizing advanced intrusion detection methods, monitoring its health status and profiling the IoT devices' behavior, as well as the collection of network information including forensic evidence; the SGA is the main link with the core platform components running at the ISP layer (only those relevant to the evidence collection process are depicted in Fig. \ref{fig.collection}). When a new device is registered, the SGA performs {\em device fingerprinting} in order to extract the device's behavioral patterns based on network flows ---\,assuming that the device is initially in a clean state. In addition, the SGA actively monitors the communication of connected devices to detect abnormal behavior by employing a lightweight IDS which transfers any suspicious traffic to the platform's back-end for {\em deep packet inspection} (DPI). Further to the above, the SGA uses {\em manufacturer's usage description} (MUD) to deliver device-focused network profiling to support accurate feature-set extraction for the anomaly detection.

More capable IoT devices, e.g. smart phones, have a {\em smart device agent} (SDA) installed that allows the direct acquisition of information (including evidence) from end-user IoT devices. The SDA operates in a more restrictive manner as it is mainly responsible for monitoring the device's usage, critical files and security ---\,firmware integrity, patching status, vulnerabilities. Information on run-time processes and the hardware resources used is regularly synchronized with the Cyber-Trust platform's back-end, and more precisely the {\em profiling service} (PS).


\subsection{Evidence collection}
\label{sec.rematt}

When suspicious network traffic and (resp. device activity) is detected by the SGA (resp. SDA), the necessary evidence is collected and sent to the ISP so as to be stored to the evDB. The evidence is comprised of IP packets (amongst other data) in the case of network attacks, whereas for device-level attacks it might include the entire device's image. At a minimum, the whole process is designed to achieve the following objectives: (a) ensure the confidentiality and integrity of forensic evidence during transmission and storage; (b) ensure that the evidence is collected from and destined to secure systems, which have established a trust relationship via an attestation protocol to authenticate the hardware/software configuration of the remote device (such as the BIOS, MBR, firmware); and (c) compute a non-repudiated proof of existence (along with other properties) of the acquired forensic evidence.

As shown in Fig. \ref{fig.collection}, the latter property is achieved by means of the CTB. The logger generates {\em evidence log events}, denoted by the {\sf EvGen} function, at the time that new evidence material is being inserted into the evidence DB, and signs these events. To achieve this step, the logger needs to have generated a key pair for use with digital signature algorithms, something that requires a {\em certificate authority} (CA) ---\,HyperLedger Fabric's CA is used for that purpose. When a new signed evidence $\ev$ is inserted in the evDB, a new identifier $\id$ is created as
\begin{equation*}
\id = \Hash \bigl( \ev \,||\, \nonce \bigr)
\end{equation*}
where the value $\nonce$ is chosen uniformly at random to ensure the uniqueness of the evidence's identifier. Note that $\id$ serves the purpose of the  signed evidence log event's integrity proof that can be verified by means of a cryptographic hash function; the evidence identifier, and the $\nonce$ used, are also stored in the evDB along with the actual data.

After computing the integrity proofs of the signed evidence log events, each proof is written to the CTB through a series of transactions, which is denoted by the function {\sf TxGen}, for subsequent generation of the next block in CTB blockchain. The blockchain explorer can then be used for retrieving the immutable record of integrity proofs on the blockchain and validate forensic evidences' properties.


\section{Forensic evidence blockchain}
\label{sec.ctb}

In the course of digital forensic investigations, the evidence examination needs to be carried out by authenticated entities, while ensuring privacy requirements. Due to this fact, only the forensic evidences' metadata are stored in the CTB, which is a permissioned distributed ledger build on HyperLedger Fabric, to provide auditing and integrity services on evidence gathered from a smart home environment. To realize the CoC and allow the entities involved to access the digital evidence, information about the chronological history of handling the evidence has to be recorded. The authenticated entities that may obtain the ownership of a forensic evidence, issue new transactions and create blocks (that contain change of ownership information), are classified as (also referred to as participants):
\begin{itemize}
\item {\em Internet service provider.} Collects the evidence regarding a security incident from the smart home environment as descibed in Section \ref{sec.rematt}. As the creator of the evidence, only the ISP is able to permanently delete it, regardless who the current owner is.
 
\item {\em Law enforcement agency.} Can access the evidentiary data about a particular $\id$, IoT device, or attack that are stored in the CTB when conducting an investigation. LEAs can also issue new transactions to transfer ownership.
 
\item {\em Prosecutor.} Considered to be the final owner of the digital forensic evidence in the course of an investigation.
\end{itemize}

\begin{figure}[t]
\centering
\includegraphics[width=\linewidth]{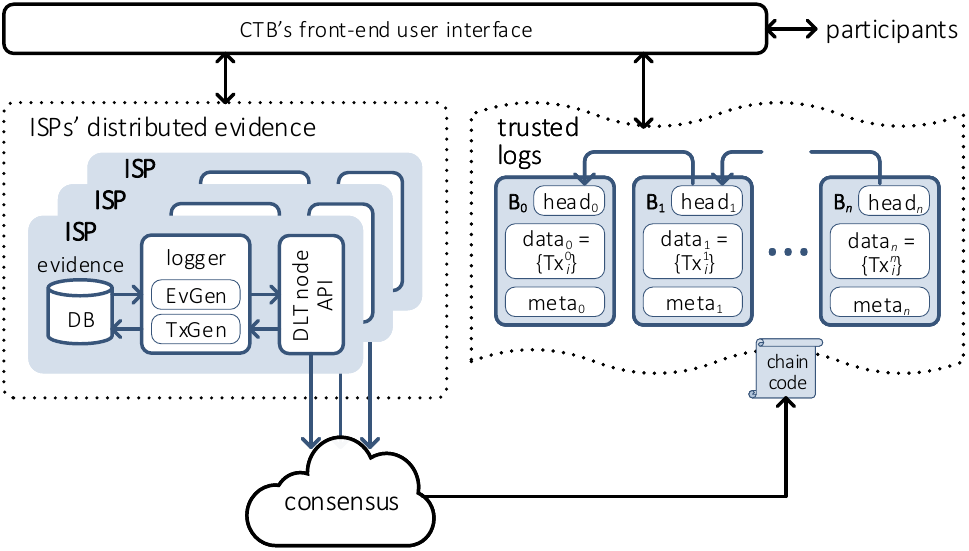}
\caption{High-level architecture of Cyber-Trust's blockchain.}
\label{fig.ctb}
\end{figure}

\noindent
In the high-level architecture of CTB, that is illustrated in Fig. \ref{fig.ctb}, the transactions stored are about the actions performed by the involved entities and also record the ownership transfer of the digital evidence from the moment of its collection until it reaches the prosecutor. The CTB is comprised of the following core components: (a) the front-end {\em user interface} (UI), (b) the blockchain node, (c) the trusted transaction logs, and (d) the forensic evidence DB. More precisely: 
\begin{itemize}
\item {\em Front-end UI.} Interface allowing the participants to view, invoke, or query blocks, transactions, chaincodes, etc., in the CTB; it is based on Fabric's blockchain explorer.

\item {\em Blockchain node.} This component ensures that authorized participants can communicate with the CTB network.
 
\item {\em Trusted logs.} Implements the blockchain and stores the historical record of facts about when evidence was created and how its ownership was transferred from one entity to another so as to arrive at the current system state.

\item {\em Forensic evidence DBs.} The off-chain databases, in which the current owner of an evidence has access to, where the raw evidentiary material is stored.
\end{itemize}
Note that there are several forensic evidence DBs, one for each ISP, and therefore, upon request of a particular evidence, the front-end UI delegates the request for access to the appropriate ISP. The design of the CTB provides  main function allowing the participants to create, transfer, erase or view the evidences stored in the evidence DB. Each function, if properly invoked, issues and broadcasts a new transaction to the network.

\setlength\partopsep{4pt}
\begin{description}[\setleftmargin{0pt}\setlabelphantom{xx}]
\setlength\itemsep{2pt}
\item [$\CreateEvidence(\id,\dsc).$] This function submits a new block to the CTB with the identifier $\id$ and the description $\dsc$ of the new evidence as input. The function's role is not just to create a new evidence, but also checks if an evidence with the same $\id$ has already been created. Another functionality is to set the first owner of the evidence, which by default is evidence's creator (i.e. the ISP).

\item [$\GetEvidence(\id).$] Given as input an evidence identifier, the function displays\,/\,retrieves the evidence after having first checked that the evidence indeed exists and the requesting participant is its current owner.

\item [$\EraseEvidence(\id).$] The function checks if the evidence with identifier $\id$ has already been stored in the CTB and if the invoking participant is the ISP that created the evidence. It is evident that forensic evidences' metadata cannot actually be erased from the CTB, as this would imply that the entire blockchain would have to be reformed. The function just deletes the evidence from the evDB and then issues a new transaction declaring that the evidence no longer exists.

\item [$\TransferOwnership(\id,\own).$] Given an evidence identifier $\id$ and a participant address $\own$, the function checks various conditions. First, the evidence must exist in the CTB and the participant invoking the function has to be the current owner of the evidence. Then, the function checks if $\own$, where the evidence will be transferred to, is authorized to access the evidence. If all conditions are true, the function transfers ownership of the evidence to the new owner $\own$, and the address of the new owner is added to the CTB.
\end{description}

\noindent
New evidence is defined as a transaction having the following metadata: the evidence identifier $\id$, the address $\creator$ of the ISP having collected the evidence, the description $\dsc$ of the security incident (initialized by the $\creator$, and later updated by other participants) and a timestamp $\tm$ of its occurrence, the current $\own$ (resp. previous $\own'$) owner of the evidence, the type ($\type$) of the attacked IoT device, as well as, the list of time records $\{\tau_i\}_{i=1,2,\ldots}$ that each owner had the evidence at his possession. The form of each transaction stored in the CTB is the following
\begin{equation*}
\Tx = \id \,||\, \creator \,||\, \dsc \,||\, \tm \,||\, \own \,||\, \own' \,||\, \type \,||\, \tau_i \,.
\end{equation*}
Let us note that, in the context of HyperLedger Fabric, only a transaction's  {\em proposal} field is shown above, which encodes the input parameters to the chaincode for creating the proposed ledger update; trivial fields, such as a transaction's header and signature, are omitted for simplicity. Since the security of CTB is of utmost importance, a number of fundamental properties need to hold \cite{Kiayias16}, the analysis of which is outside the scope of this work, such as persistence, liveness, chain quality property, and common prefix property. If all true, they considerably limit the ability of adversaries to alter CTB evidentiary metadata.


\section{Conclusions}
\label{sec.concl}

Cyber-Trust platform relies on advanced intrusion detection tools to identify malicious activities and enhance the security of IoT environments by inspecting compromised devices and collecting forensic evidence so as to determine the source of cyber-attacks. The evidentiary information is safely stored as raw data in an off-chain database, while the hashes and metadata of the evidence are stored on the blockchain. The~CTB is a permissioned distributed ledger, which is build on top of HyperLedger Fabric. Cyber-Trust's blockchain-based solution dematerializes the CoC process of recording and preserving a chronological history of digital evidences.


\bibliographystyle{IEEEtran}
\bibliography{Paper}
\end{document}